\documentclass[twocolumn,floatfix,nofootinbib,
tightenlines,nobibnotes,aip,apl,10pt]{revtex4-1}

\usepackage{natbib}
\usepackage{graphicx}
\usepackage{upgreek}

\usepackage{graphicx}
\usepackage{amssymb}
\usepackage{amsmath}
\usepackage{textcomp}

\begin{document}
\title{Coherent-State Storage and Retrieval Between Superconducting Cavities Using Parametric Frequency Conversion}

\author{A. J. Sirois}
\affiliation{National Institute of Standard and Technology,  Boulder, CO 80305 USA}
\affiliation{University of Colorado - Boulder, CO 80309 USA}
\author{M. A. Castellanos-Beltran}
\affiliation{National Institute of Standard and Technology,  Boulder, CO 80305 USA}
\author{M. P. DeFeo}
\affiliation{National Institute of Standard and Technology,  Boulder, CO 80305 USA}
\author{L. Ranzani}
\affiliation{National Institute of Standard and Technology,  Boulder, CO 80305 USA}
\author{F. Q. Lecocq}
\affiliation{National Institute of Standard and Technology,  Boulder, CO 80305 USA}
\author{R. W. Simmonds}
\affiliation{National Institute of Standard and Technology,  Boulder, CO 80305 USA}
\author{J. D. Teufel}
\affiliation{National Institute of Standard and Technology,  Boulder, CO 80305 USA}
\author{J. Aumentado}
\affiliation{National Institute of Standard and Technology,  Boulder, CO 80305 USA}

\date{\today}

\begin{abstract}

In superconducting quantum information, machined aluminum superconducting cavities have proven to be a well-controlled, low-dissipation electromagnetic environment for quantum circuits such as qubits. They can possess large internal quality factors, $Q_{int}>10^8$, and present the possibility of storing quantum information for times far exceeding those of microfabricated circuits. However, in order to be useful as a storage element, these cavities require a fast ``read/write'' mechanism--- in other words, they require tunable coupling between other systems of interest such as other cavity modes and qubits, as well as any associated readout hardware. In this work, we demonstrate these qualities in a simple dual cavity architecture in which a low-Q ``readout'' mode is parametrically coupled to a high-Q ``storage'' mode, allowing us to store and retrieve classical information. Specifically, we employ a flux-driven Josephson junction-based coupling scheme to controllably swap coherent states between two cavities, demonstrating full, sequenced control over the coupling rates between modes.

\end{abstract}


\maketitle

Quantum optics experiments which explore atomic or optomechanical physics often rely on parametric or ``sideband'' interactions to achieve strong coupling between modes at different frequencies.\cite{ExpQuant} By contrast, quantum optics experiments based on superconducting circuits most often employ resonant interactions to achieve strong coupling.\cite{WilhelmClarke,GenSinglePhoton,GenFock} While these techniques are reliable and well-understood, they are difficult to scale to more complex systems as they require either more physical wiring or sophisticated multiplexing of DC shift pulses. One way to avoid these difficulties is to engineer non-resonant parametric coupling in these systems. For instance, these techniques have been used to convert single microwave photons between different frequencies of a coplanar waveguide resonator,\cite{EvaAndFrancois} and between a qubit and a linear mode.\cite{plourde,allman2014tunable} They have also been used to exchange coherent states between propagating modes at different frequencies,\cite{AbdoYale} and between a cavity mode and the internal modes of a Josephson Ring Modulator.\cite{Huard} Here we extend these results by demonstrating parametric frequency conversion of coherent states between two spatially-distinct, three-dimensional cavity resonators. The use of non-resonant interactions implies the use of only static biasing fields and all-microwave frequency control pulses to switch the interactions on and off. Moreover, it allows for additional control over the coupling phase, as well as the ability to sequence coupling between any arbitrary pair of modes. This general approach possesses the potential to reduce wire count in quantum computing architectures since coupling in multi-mode systems can be frequency-multiplexed. Finally, this work demonstrates a methodology which may be useful in novel approaches to quantum computing that use linear modes for state storage.\cite{DynProtectedCats}

\begin{figure}
\includegraphics[scale=0.87]{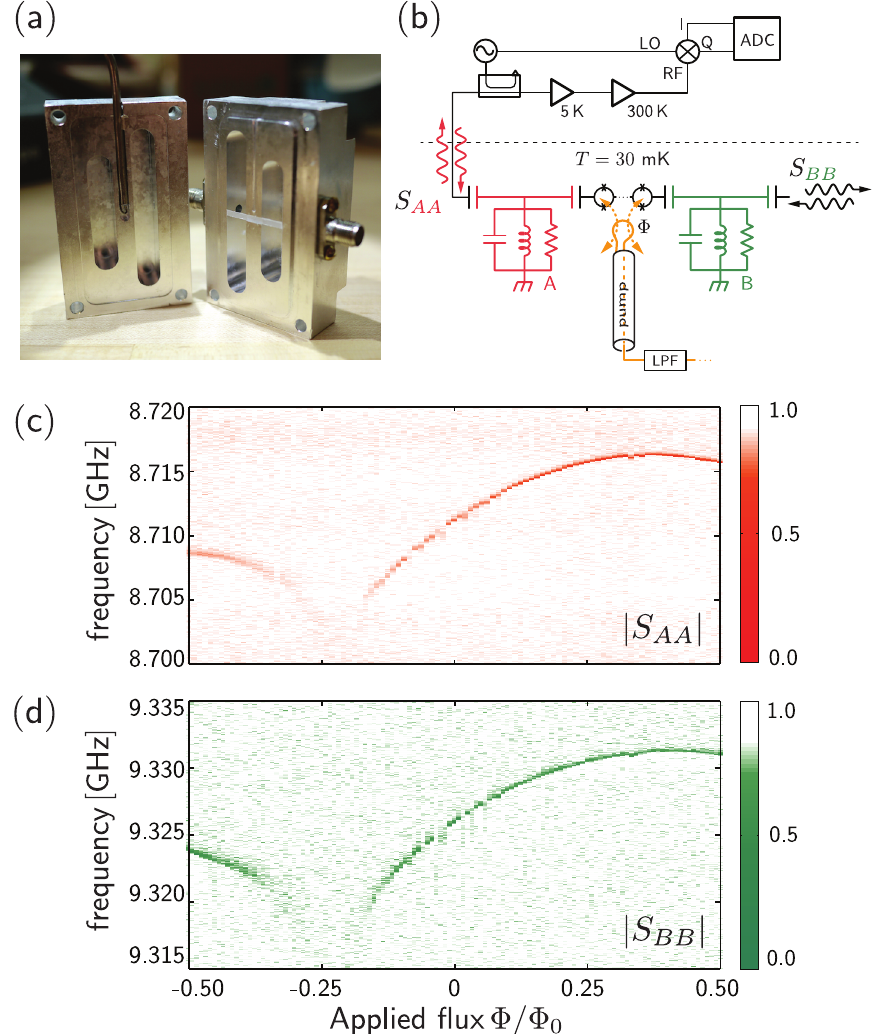}
\caption{(a) Bottom half of the aluminum dual-cavity box showing the two cavity volumes, sapphire coupling chip, and SMA connectors. The pump line coax is mounted in the top half (\emph{not shown}). (b) Schematic of the general measurement scheme.(c-d) Reflection coefficients of the readout and storage modes as a function of flux bias on the coupling chip SQUIDs. There is an offset in the modulation curve due to stray flux in the coupler.}
\label{fig:devicefigure}
\end{figure}

In this work we modulate a reactance which couples the electromagnetic fields of two physically separated aluminum cavities (see Figure\,\ref{fig:devicefigure}(a--b)). The dynamics of such a system can be encapsulated by the coupled mode equations of motion,\cite{Louisell} 

\begin{eqnarray}
\frac{da(t)}{dt} &= -i (\omega_{A} - i\frac{\gamma_A}{2})a(t) - i g_{P}(t) e^{i(\omega_{p}t + \phi_{P})} b(t)\\
\frac{db(t)}{dt} &= -i (\omega_{B} - i\frac{\gamma_B}{2})b(t) - i g_{P}(t) e^{-i(\omega_{p}t + \phi_{P})} a(t)
\label{eq:coupledmodeEOM}
\end{eqnarray}
where $a(t)$ and $b(t)$ are the complex mode amplitudes of the electromagnetic field in the readout (``A'') and storage (``B'') cavities (normalized to the square root of the photon energies, $\sqrt{\hbar\omega_{A/B}}$),  $\omega_{A/B}$ are the natural frequencies of the readout and storage cavities, and $\gamma_{A/B}$ are the corresponding total dissipation rates. In the highly off-resonant regime where the mode separation $|\omega_A - \omega_B|$ is much greater than the static coupling and dissipation rates, energy cannot be exchanged between modes residing within these cavities. However, if the coupling term is modulated at the difference frequency, $\omega_P = |\omega_A - \omega_B|$, such modes can resonantly couple in a mean rotating frame, transferring energy in both directions. This process, parametric frequency conversion, can be turned on and off by modulating the coupling envelope amplitude $g_P(t)$. In addition, it provides an extra degree of control over the swap process through the pump phase $\phi_P$.

\begin{figure*}
\includegraphics[scale = 1.0]{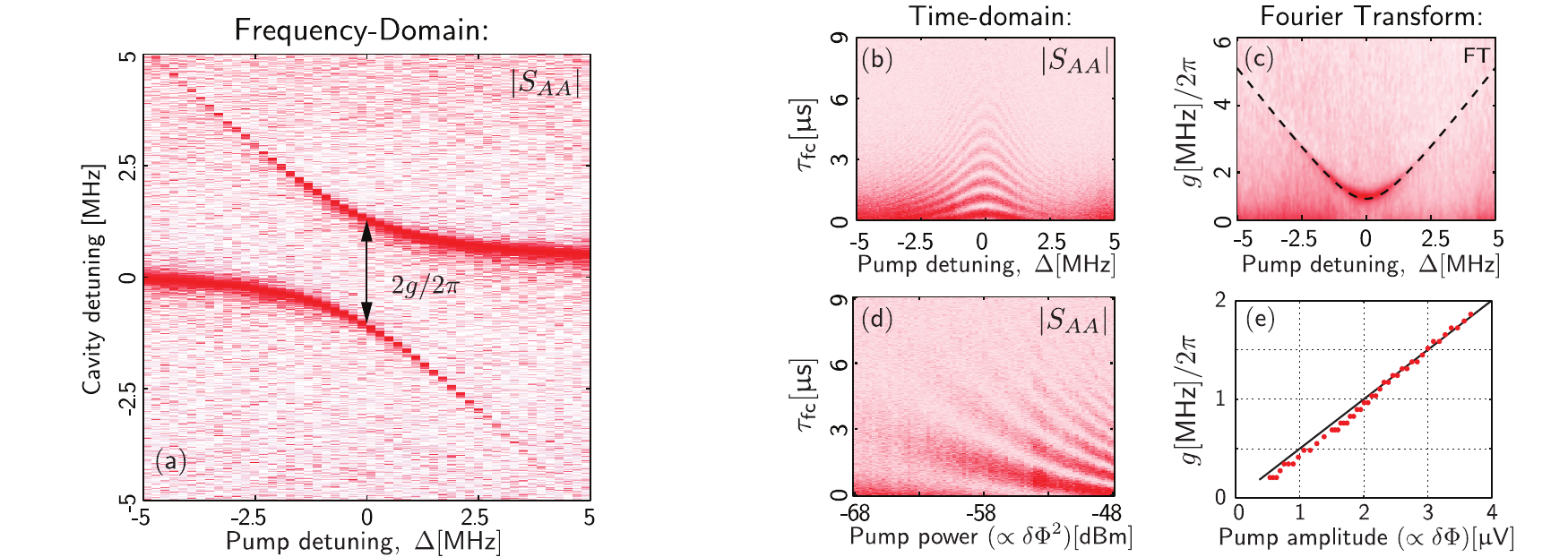}
\caption{(a) Steady-state frequency-domain response of cavity A to a moderate (-52 dBm) pump tone sweeping through $\omega_{p}$ near $|\omega_{A} - \omega_{B}|$ showing the expected parametrically-induced normal-mode splitting of $2g_P=2\pi 2.4$ MHz. (b,c) Time-domain response and Fourier transfrom of the energy in cavity A with a swap tone at a fixed power applied after loading a coherent state into cavity A. (d) Time-domain response of the retrieved cavity A energy as a function of pump power ($\Delta=0$). (e) The swap rate $g_P$ derived from the Fourier transformed data in (d). In (b,d) $\tau_{fc}$ is the pulse length of the swap tone.}
\label{fig:CWresponse}
\end{figure*}

Our two cavities are milled out of a split aluminum block (Figure\,\ref{fig:devicefigure}(a)), where each cavity is similar in construction to previously published long-lived cavity resonator designs.\cite{Yale10ms} The readout (A-mode) cavity (6.35\,mm $\times$ 19.05\,mm $\times$\,38.10\,mm, $\omega_{A} = 2\pi$\,$8.70$\,GHz (TE101 mode)) is intended to couple strongly to a transmission line routed into our measurement chain, while the storage (B-mode) cavity (6.35\,mm\,$\times$\,19.05\,mm\,$\times$\,25.40\,mm, $\omega_{B} = 2\pi$\,$9.33$\,GHz (TE101 mode)) is designed to minimize external dissipation, although we intentionally increase this coupling when performing initial characterization measurements.

The reactance that provides the modulated coupling in our system is implemented as an array of ten superconducting quantum interference devices or SQUIDs (critical current $I_c = 1.5\,\upmu$A, loop area = $50\,\upmu$m$\,\times\,50\,\upmu$m, total array length $\simeq 8\,$mm) patterned onto a sapphire chip which is placed in a tunnel connecting the two cavities. The SQUID array couples into their electromagnetic fields via long patterned probes (15 $\upmu$m wide, 90 nm thick) extending 1.5\,mm into each cavity volume.  The SQUIDs act as a flux-tunable inductance which is shunted by stray capacitance to the aluminum box, forming a tunable self-resonance with a maximum natural frequency $\omega_{C, max} \simeq 2\pi 7.7\,$GHz. Since $\omega_{C,max}<\omega_A,\omega_B$, the coupler actually behaves as a flux-modulated effective capacitance coupling the two cavities. As we show below, this configuration is sufficient to create parametric coupling between resonator modes in this system. Both AC and DC flux are driven by a loop-terminated niobium coaxial cable (the ``flux pump line'') placed near the SQUID array, with a mutual inductance to each SQUID $M\simeq$\,0.05 pH. 

All measurements are performed in a dilution refrigerator at 30 millikelvin, using standard cryogenic filtering and microwave measurement techniques (Figure\,\ref{fig:devicefigure}(b)). In our initial measurements we characterize the sensitivity of the cavity frequencies to DC magnetic flux applied to the SQUID array. In Figure\,\ref{fig:devicefigure}(c--d) we observe the natural frequencies, $\omega_{A/B}$, modulating by several megahertz, demonstrating that the coupler element participates significantly in the individual cavity mode impedances-- a basic requirement for dual-mode parametric coupling. From this we can estimate the parametric coupling rate amplitude,\cite{EvaAndFrancois}
\begin{equation}
g_P = \frac{\delta\Phi}{4}\sqrt{\frac{\partial \omega_A}{\partial\Phi}\frac{\partial \omega_B}{\partial\Phi}},
\label{eq:gp}
\end{equation}
where $\delta\Phi$ is the amplitude of the pump flux modulation and the partial derivatives are the first order terms in a Taylor expansion of the flux modulation curves shown in Figure~\ref{fig:devicefigure}(c-d) expanded around a given DC flux bias point. We note that in these initial spectroscopic measurements, coupling to the pump line and its 50\,$\Omega$ environment created unwanted dissipation, limiting the total quality factor of the A and B cavities. In the remaining measurements we mitigated this dissipation channel by placing a commercial inline low pass filter ($f_{LPF}\simeq 1\,$GHz) to isolate the cavity mode frequencies while maintaining access to the lower-frequency pump tones. In addition, we removed the coupling between cavity B and its associated 50\,$\Omega$ transmission line. The measured internal quality factor is $Q_A^{int}\simeq 900,000$, while the external quality factor is $Q_A^{ext}\simeq50,000$, set by the length of the microwave connector pin extending into the cavity volume. The corresponding total dissipation for this mode yields $\gamma_A \simeq 2\pi170$\,kHz and $T_1^A \simeq 0.94\,\upmu$s. 

In Figure~\ref{fig:CWresponse}(a) we modulate the pump flux, $\delta\Phi \simeq 0.2 \Phi_0$ ($P_P= -52$\,dBm), and observe splitting in the continuous wave reflected response of cavity $A$ as a function of pump detuning $\Delta \equiv\omega_P - |\omega_A - \omega_B|$. This indicates continuous energy exchange with a mode at $\omega_A + \omega_P = 2\pi\, 9.33$\,GHz, \emph{i.e.}, the expected cavity B natural frequency. The observed splitting size, $2g_P=2\pi\,2.4$\,MHz is consistent with the prediction from equation \ref{eq:gp} and our flux bias. We are able to increase the coupling rate up to $g_P \simeq 2\pi\,3.5$\,MHz before the system begins to heat from excess pump power dissipation at the mixing chamber cooling stage.

These continuous wave measurements show that strong coupling is possible between our two cavity modes. To demonstrate this in the time domain we inject a coherent state into mode $A$ (such that the photon number population is $\bar{n}_A=10$), pulse the pump drive for a fixed length of time, $\tau_{fc}$, swapping some fraction of the energy between $A$ and $B$, and then integrate the power that leaked from the $A$ port into our amplification chain. Figure~\ref{fig:CWresponse}(b) shows the (uncalibrated) leaked energy from $A$ as a function of both pump drive detuning and length of the pump envelope. These data show Rabi-like oscillations as a function of swap time and pump frequency detuning as expected for coupled classical linear oscillators.\cite{Louisell,frimmer2014classical} In Figure~\ref{fig:CWresponse}(c) we show the Fourier transform of these oscillations demonstrating a $\sqrt{\Delta^2 + g_P^2}$ detuning dependence.\cite{EvaAndFrancois} In a similar manner we fix the pump frequency ($\Delta = 0$) and measure the pump power dependence of these swap oscillations (Figure~\ref{fig:CWresponse}(d)). The resulting swapping rate, $g_P$, follows the expected linear dependence on pump amplitude up to a few megahertz (Figure~\ref{fig:CWresponse}(e)). 

\begin{figure}
\includegraphics[scale = 0.88]{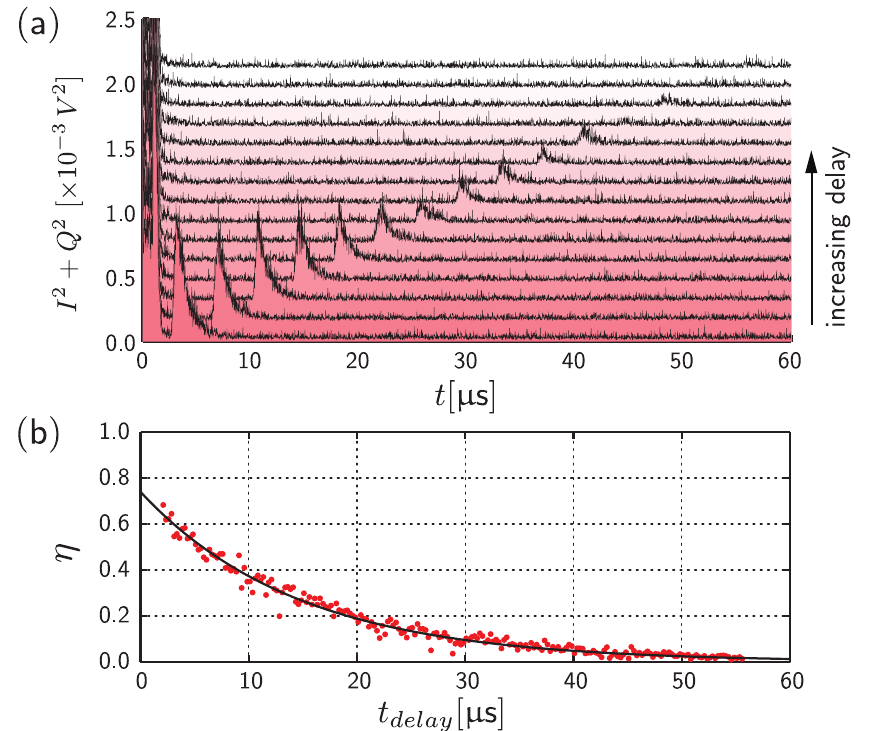}
\caption{(a) Magnitude ($I^2 + Q^2$) \emph{vs.} time of the retrieved state signal leaking from Cavity A for several delay times ($t_{delay} = 1$--55$\upmu$s, offset for clarity). (b) Retrieval efficiency $\eta$ (red dots, derived from the integrated IQ traces) \emph{vs.} delay time. The fit (solid black) yields a decay time $\tau_B$  = 14.9\,$\upmu$s ($\gg\tau_A = 0.93\,\upmu$s).}
\label{fig:StoreRetrieve}
\end{figure}

The swap data shown in Figure~\ref{fig:CWresponse}(b) provide the necessary pulse-length calibration for a full state swap, analogous to the $\pi$-pulse calibration one would do when driving to the excited state in a qubit. Fixing the pump frequency, power, and envelope time accordingly ($\omega_P = 2\pi\,632.5$\,MHz; $P_P$ = -52\,dBm, $t_{swap} = $0.6\,$\upmu$s), we can perform a storage and retrieval measurement sequence. In this sequence, we inject a coherent state into readout mode $A$, frequency convert/swap it into storage mode $B$, wait a delay time $t_{delay}$, then swap it back to $A$ where it leaks out rapidly into the amplification chain. Typical time domain data for the leaked power are shown in Figure~\ref{fig:StoreRetrieve}(a). Again, we integrate these traces to provide a measure of the total energy retrieved. Plotting this as a function of delay time yields a $B$-mode decay time $T_1^B = $\,14.9\,$\upmu$s (Figure\,\ref{fig:StoreRetrieve}(b)). This time far exceeds the measured readout mode decay time, $T_1^A$ noted earlier. 

The integrated energy retrieved (after the shortest storage time) can be compared to the energy stored in cavity A with the swap sequence disabled to derive an efficiency measure, $\eta$, of the swap process -- this yields $\eta=74.7\%$. We note that this efficiency is dominated by the decay of cavity A during the swap pulses. If we account for the energy loss from the decay of cavity A during the swap, the swapping efficiency increases to $\eta'=99.6\%$, indicating that the swap process itself does not add significant loss.

\begin{figure}
\includegraphics[scale = 0.87]{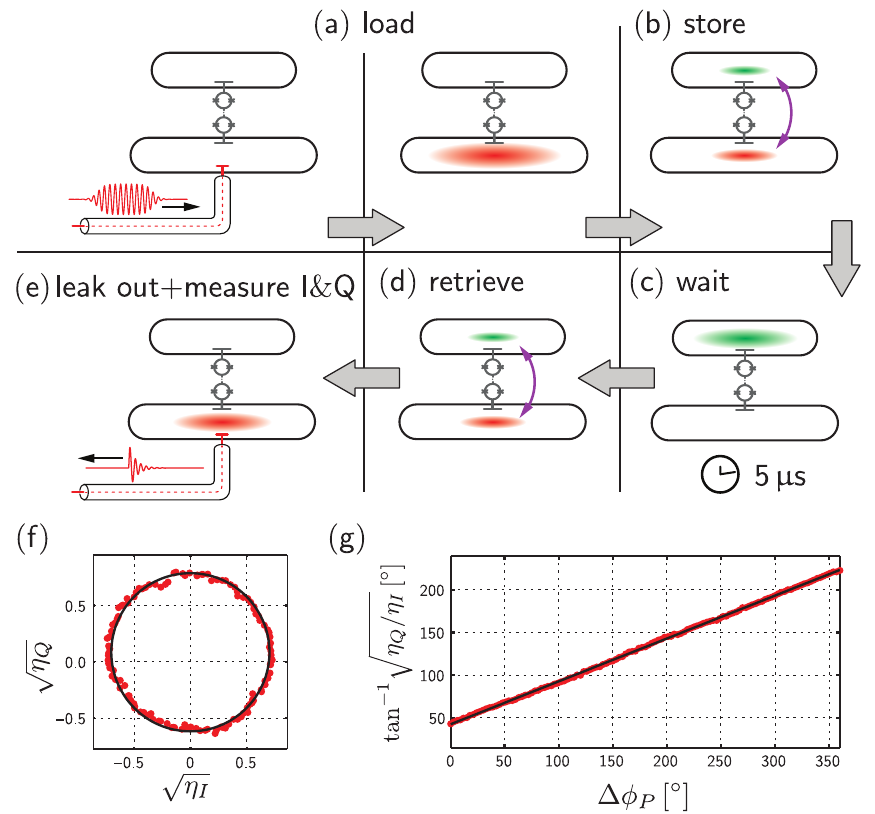}
\caption{State initialization and measurement sequence. (a) a coherent state is loaded into cavity A with a 20 $\upmu$s pulse; (b) the coherent state is immediately swapped to cavity B with a 0.6$\upmu$s pulse applied to the pump at the difference frequency, $\omega_{p}$; (c) the coherent state is stored in mode B for 5 $\upmu$s; (d) the same pulse as in (b) is applied again, but with a phase shift $\Delta\phi_P$ relative to the first swap-pulse; (e) the final state in mode A is then read out and demodulated. (f) Efficiency of the $I$ and $Q$ integrated signals over a $2\pi$ range of relative phase shifts. The corresponding data (red dots) and its fit (solid black) are scaled such that they encompass an area $=\pi\eta$, where $\eta = 0.49$, the state retrieval efficiency at $5\,\upmu$s delay. (g) Phase of the retrieved integrated signal \emph{vs.} relative phase shift between the storage and retrieval frequency conversion drives. The fit slope (solid black) is 1.
}
\label{fig:IQSequence}
\end{figure}

We complete our characterization of the parametric swap process by demonstrating control over the coupling phase. As before, we perform a storage and retrieval sequence (see Figure~\ref{fig:IQSequence}(a-e)), but fix the pump power, duration, and delay time, changing only the pump phase on the retrieval pulse.  We can clearly see that the pump phase is imprinted on the resulting retrieved state that leaked from the A-mode into the amplification chain, Figure~\ref{fig:IQSequence}(f). In Figure~\ref{fig:IQSequence}(g) we show the phase angle of the retrieved signals as a function of relative pump phase. This ability to easily control the phase in the swap process using vector modulation sets parametric frequency conversion apart from resonant coupling and opens other possibilities for manipulating the state of dual-mode systems.

In summary, we have demonstrated the storage and retrieval of coherent microwave states in a dual 3D cavity system using a controllable parametric modulation scheme. This approach may be extended to systems with higher coherence time systems,\cite{Yale10ms,creedon2011high} and may provide an essential resource for a recently proposed fault-tolerant quantum information architecture based on linear modes.\cite{DynProtectedCats}

\bibliographystyle{unsrt}
\bibliography{StoreAndRetrieveBib}

\end{document}